\documentclass[a4paper,11pt]{article}
\pdfoutput=1 

\usepackage{jcappub} 

\usepackage[T1]{fontenc} 
\usepackage[utf8]{inputenc}
\usepackage{amsfonts}
\usepackage{lmodern}
\usepackage{amsthm,graphicx}
\usepackage{epsfig}
\usepackage{ulem}
\usepackage{latexsym, amssymb} 
\usepackage{amsmath}

 \usepackage{xcolor}
 \newcounter{attnctr} 

\title{\boldmath Unveiling acoustic physics of the CMB using nonparametric estimation of the temperature angular power spectrum for Planck}

\author[a]{Amir Aghamousa,}
\author[a,b]{Arman Shafieloo,}
\author[c]{Mihir Arjunwadkar}
\author[d]{and Tarun Souradeep}

\affiliation[a]{Asia Pacific Center for Theoretical Physics,\\Pohang, Gyeongbuk 790-784, Korea}
\affiliation[b]{Department of Physics,\\POSTECH, Pohang, Gyeongbuk 790-784, Korea}
\affiliation[c]{Centre for Modeling and Simulation,\\Savitribai Phule Pune University, Ganeshkhind, Pune 411 007, India}
\affiliation[d]{Inter-University Centre for Astronomy and Astrophysics,\\Post Bag 4, Ganeshkhind, Pune 411 007, India}

\emailAdd{amir@apctp.org}
\emailAdd{arman@apctp.org}
\emailAdd{mihir@cms.unipune.ac.in}
\emailAdd{tarun@iucaa.ernet.in}

\abstract{Estimation of the angular power spectrum is one of the important steps in Cosmic Microwave Background (CMB) data analysis.
Here, we present a nonparametric estimate of the temperature angular power spectrum for the Planck 2013 CMB data.
The method implemented in this work is model-independent, and allows the data, rather than the model, to dictate the fit.
Since one of the main targets of our analysis is to test the consistency of the $\Lambda$CDM model with Planck 2013 data, we use the nuisance parameters associated with the best-fit $\Lambda$CDM angular power spectrum to remove foreground contributions from the data at multipoles $\ell \geq50$.
We thus obtain a combined angular power spectrum data set together with the full covariance matrix, appropriately weighted over frequency channels.
Our subsequent nonparametric analysis resolves six peaks (and five dips) up to $\ell \sim1850$ in the temperature angular power spectrum.
We present uncertainties in the peak/dip locations and heights at the $95\%$ confidence level. We further show how these reflect the harmonicity of acoustic peaks, and can be used for acoustic scale estimation.
Based on this nonparametric formalism, we found the best-fit $\Lambda$CDM model to be at $36\%$ confidence distance from the center of the nonparametric confidence set -- this is considerably larger than the confidence distance (9\%) derived earlier from a similar analysis of the WMAP 7-year data.
Another interesting result of our analysis is that at low multipoles, the Planck data do not suggest any upturn, contrary to the expectation based on the integrated Sachs-Wolfe contribution in the best-fit $\Lambda$CDM cosmology.}

\begin{document}
\maketitle
\flushbottom

\section{Introduction}
\label{sec:intro}
Anisotropy measurements for the Cosmic Microwave Background (CMB)
provide high precision information about the Universe, largely encoded in its temperature angular power spectrum.
Especially, the shape of this angular power spectrum is a sensitive measure of the cosmological parameters and initial conditions of a `standard' homogeneous and isotropic Universe.
Therefore, it is important to gauge how realistic and reasonable the estimated angular power spectrum is before drawing conclusions about the Universe.

Conventional methods used by cosmologists are based on assuming a specific family of cosmological models with finite number of free parameters to infer the underlying theoretical angular power spectrum.
On the other hand, nonparametric methods do not assume any specific model form; they attempt to estimate the true but unknown angular power spectrum with minimum possible assumptions \citep{Aghamousa2012}.

The specific nonparametric method employed in this work estimates the angular power spectrum along with a high-dimensional confidence set (often referred to as \textit{confidence ball} because it is ellipoidal by construction).
The confidence ball also allows the inference of key features such as uncertainties on locations and heights of peaks and dips of the inferred angular power spectrum that are important for capturing and verifying the fundamental physical processes of the underlying cosmology.
It can also be used for validating different cosmological models, based on the data.
This nonparametric methodology was first introduced in a series of papers
\citep{Beran1996, BD1998, Beran2000, Beran2000b}, which was then generalized for the case of known noise covariance matrix, and used in CMB data analysis by \citep{GMN+2004} and \citep{BSM+2007}.
The method was further adopted with important improvements for the analysis of the WMAP 1, 3, 5, and 7-year data sets  \citep{Aghamousa2012}, and also used \citep{Aghamousa2014} to forecast, using simulated Planck-like data, the temperature and polarization angular power spectra expected from the Planck mission \citep{TPC2006}.

In this paper, we present the nonparametric analysis of the recently released Planck temperature angular power spectrum data \citep{Planck2013I}.
First, we obtain a combined angular power spectrum data and its covariance matrix using the angular power spectra data at different frequency channels (Sec.\ $\S$~\ref{subsec:data}).
Using this combined angular power spectrum, we estimate the nonparametric fits, as described in $\S$~\ref{subsec:nonparam-spectrum}.
In $\S$~\ref{subsec:quality}, we present an assessment of the quality of our nonparametrically estimated spectrum.
In $\S$~\ref{subsec:boxes}, we present the uncertainties on peak/dip locations and heights of the nonparametric angular power spectrum.
In $\S$~\ref{subsec:harmonicity_acoustic}, we demonstrate the harmonicity of acoustic peaks of this angular power spectrum, and present an estimate of the acoustic scale $\ell_A$ which is related to the widely-accepted acoustic physics of the standard cosmological model.
In $\S$~\ref{subsec:checking}, we use our nonparametric spectrum to check the consistency of the Planck best-fit $\Lambda$CDM model \citep{Planck2013XV, Planck2013XVI} with the data.
Finally, we present our conclusions in $\S$~\ref{sec:conclusions}.
\section{Data and methods}~\label{sec:data_method}
The Planck mission measured the CMB temperature anisotropies over the whole sky with higher precision than preceding experiments \citep{Planck2013I}.
It observes the sky in different frequency bands from 30 to 353 GHz, and provides the temperature angular power spectrum in the multipole range $2\leq \ell \leq 2500$.
For estimating the temperature angular power spectrum, which is the goal of this paper, we first construct, as described below, a single angular power spectrum data set by appropriately combining data at different frequency channels.

\subsection{Data}~\label{subsec:data}
The Planck satellite observes the CMB temperature in a wide range of channels from 30 to 353 GHz.
For $\ell<50$, where Galactic contamination has a significant contribution,
the angular power spectrum data is derived by
using all channels to remove Galactic foregrounds and give a single data set.
At the $\ell \geq50$ multipoles, the extragalactic foregrounds are relatively more significant.
The Planck likelihood code implements a model with a set of nuisance parameters to account for the effects of the foregrounds.
It calculates the likelihood value by using the foreground model and the angular power spectra in the frequency range from 100 to 217 GHz.

\begin{figure}
 \includegraphics[width=\textwidth]{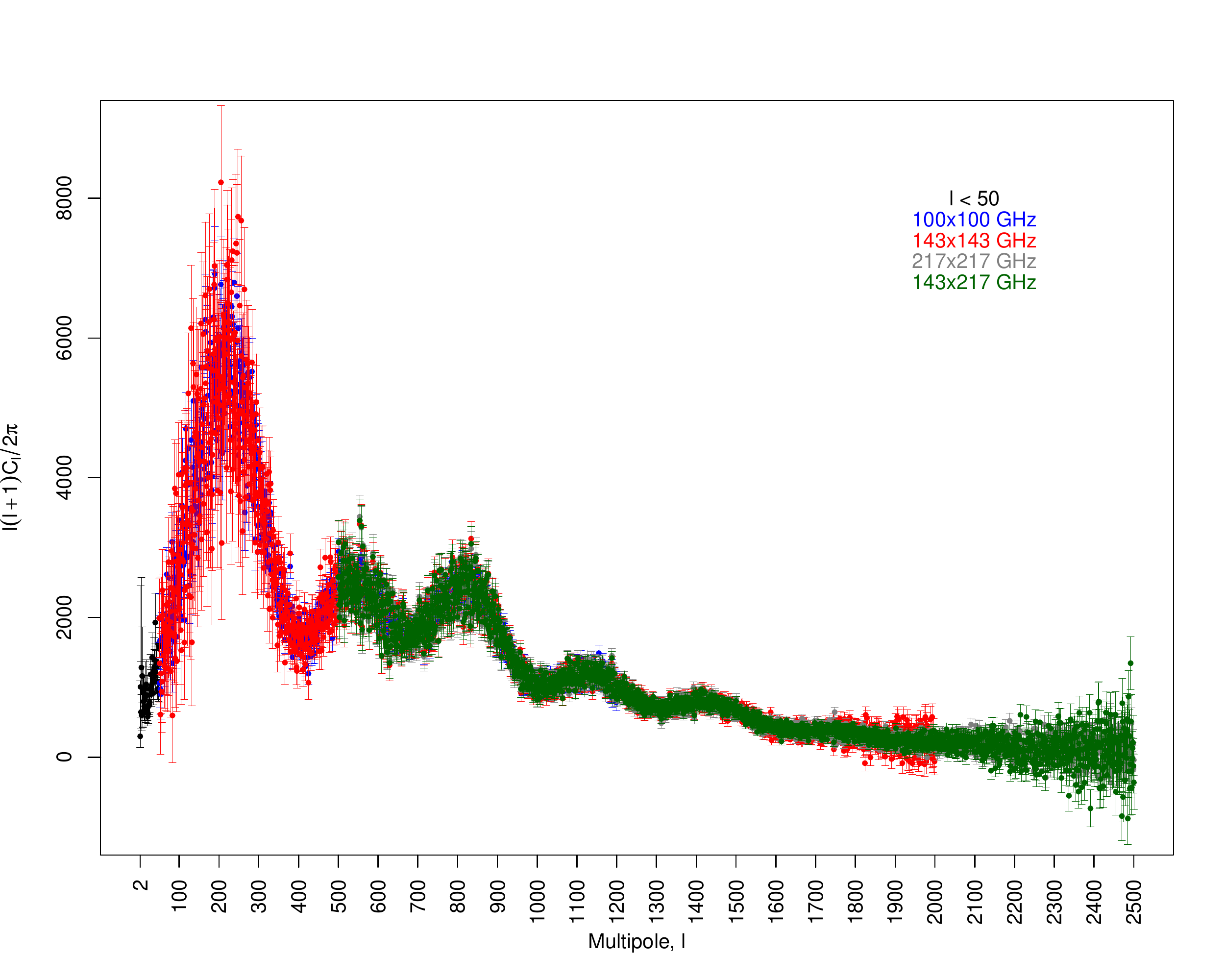}
 \caption{\label{fig:data1} The Planck angular power spectra data. The angular power spectrum at low multipoles ($\ell<50$) is obtained by using the frequency channels from 30 to 353 GHz (black points and error bars). For the higher multipoles ($\ell \ge50$), the background Planck angular power spectra in channels $100 \times 100$ GHz, $143\times143$ GHz, $217\times217$ GHz and $143\times217$ GHz (blue, red, gray and green points and error bars respectively) are obtained by using the foreground nuisance parameters associated with the best-fit $\Lambda$CDM model \citep{Planck2013XVI, Planck2013XV}.}
\end{figure}

Because we would like to compare model-independent estimate of the angular power spectrum with the Planck best-fit $\Lambda$CDM model, we borrow the nuisance parameters associated with the Planck best-fit $\Lambda$CDM model \citep{Planck2013XVI,Planck2013XV} to obtain the background angular power spectra data in the frequency range 100 to 217 GHz.
Figure~\ref{fig:data1} depicts the background angular power spectra thus obtained.


Table~\ref{tab:cross-spectra} tabulates the angular power spectra with corresponding coverage multipoles range.
As can be seen, they overlap in some ranges of multipoles.
To obtain a single data set, we calculate a weighted-average $\bar{{\cal D}_{\ell}}$ of angular power spectra ${\cal D}_{\ell}= \ell (\ell+1)C_{\ell}/2\pi$ over contributing frequency channels for each multipole $\ell$.
Individual angular power spectra are weighted inversely in proportion of their corresponding variances.
The weighted-average angular power spectrum $\bar{{\cal D}_{\ell}}$ thus has the form
\begin{eqnarray}\label{equ:weight-average1}
\bar{{\cal D}_{\ell}} &=& \frac{\sum_{ch}w^{ch}_{\ell} {\cal D}^{ch}_{\ell}}{\sum_{ch}w^{ch}_{\ell}}, \\
 w^{ch}_{\ell}&=&(\sigma^{ch}_{\ell})^{-2}, \nonumber
\end{eqnarray}
where $ch$ runs over channels in Table~\ref{tab:cross-spectra}, ${\cal D}^{ch}_{\ell}$ is the temperature angular power spectrum of the associated frequency channel (in $\mu K^2$ units) at multipole
$\ell$, $w^{ch}_{\ell}$ is the weight term, and $(\sigma^{ch}_{\ell})^{-2}$ is the inverse estimated variance of ${\cal D}^{ch}_{\ell}$.

\begin{figure}
 \includegraphics[width=\textwidth]{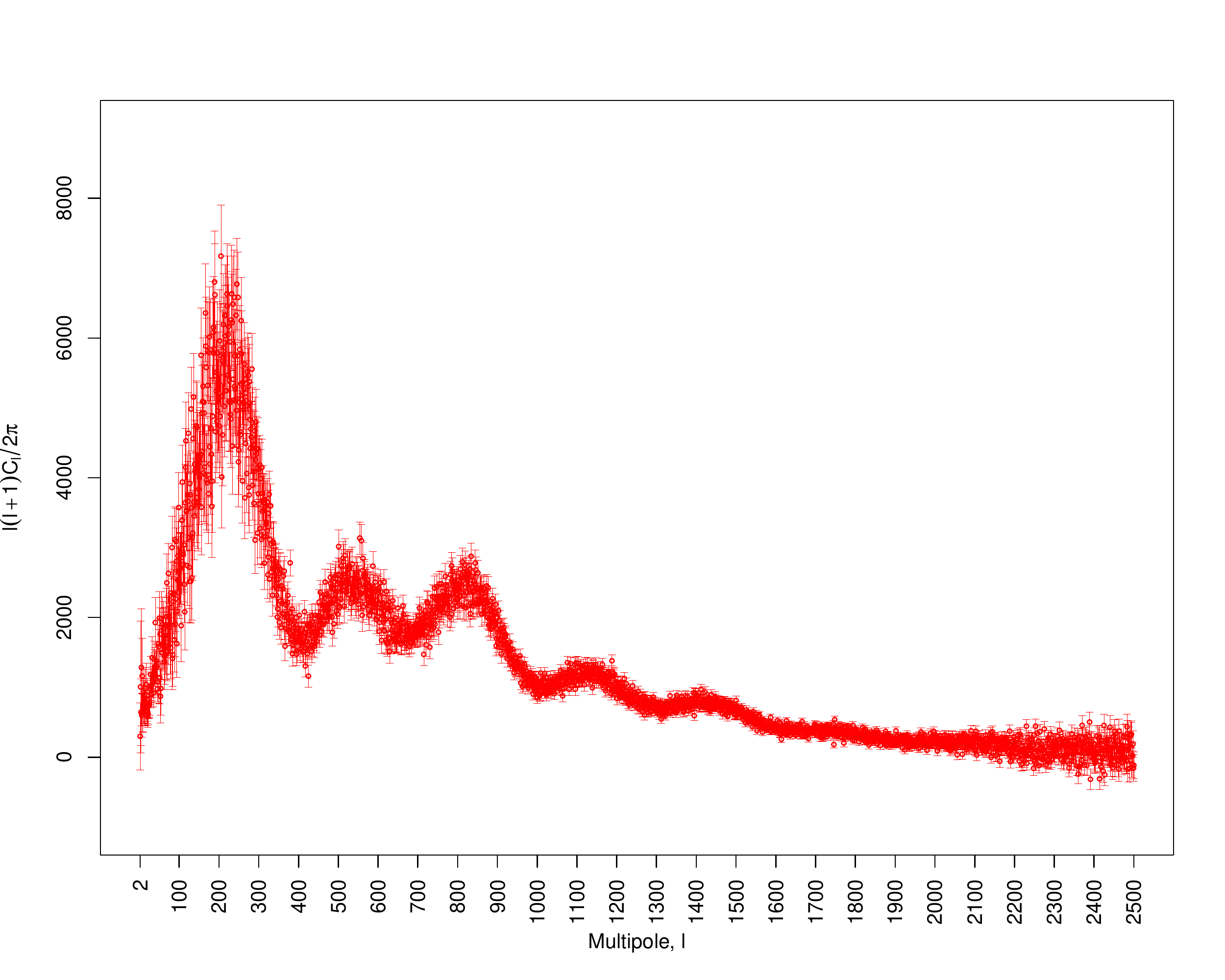}
 \caption{\label{fig:data2} The combined Planck temperature angular power spectrum data. The angular power spectrum in $\ell<50$ is obtained by using the Planck frequency channels from 30 to 353 GHz. For higher multipoles ($\ell \ge50$) the weighted-average of the background CMB angular power spectra from combination of $100 \times 100$ GHz, $143\times143$ GHz, $217\times217$ GHz and $143\times217$ GHz channels is used.}
\end{figure}

The covariance matrix of the combined data, considering all correlation terms between different spectra and multipoles, takes the form
\begin{eqnarray}\label{equ:weight-covariance}
\text{Cov}(\bar{{\cal D}_{\ell}},\bar{{\cal D}_{\ell'}}) &=& \frac{1}{\sum_{ch}w^{ch} \sum_{ch'}w^{ch'}} \quad \text{Cov}(\sum_{ch}w^{ch}_{\ell} {\cal D}^{ch}_{\ell},\sum_{ch'}w^{ch'}_{\ell'} {\cal D}^{ch'}_{\ell'}) \nonumber \\
&=& \frac{1}{\sum_{ch}w^{ch} \sum_{ch'}w^{ch'}} \sum_{ch, ch'} w^{ch}_{\ell} w^{ch'}_{\ell'} \text{Cov}( {\cal D}^{ch}_{\ell}, {\cal D}^{ch'}_{\ell'}),
\end{eqnarray}
where $ch$ and $ch'$ run over frequency channels in Table~\ref{tab:cross-spectra}, and $\text{Cov}( {\cal D}^{ch}_{\ell}, {\cal D}^{ch'}_{\ell'})$ indicates the covariance between frequency channels $ch$ and $ch'$ at  multipoles $\ell$ and $\ell'$ as computed by the Planck likelihood code \citep{Planck2013XV}.
As we can see from the above equation, this covariance matrix incorporates correlations between all the contributing frequency channels.
Figure~\ref{fig:data2} shows the weighted-average angular power spectrum $\bar{{\cal D}_{\ell}}$ for $2\leq \ell \leq2500$, with error bars based on the diagonal terms of the calculated covariance matrix (Equation~\ref{equ:weight-covariance}).

\begin{table}[!htb]
\begin{center}
\vspace{6pt}
\begin{tabular}{| l | c |}
\hline
Spectrum  &  Multipole range\\
\hline
$100\times100$  &  50-1200   \\
$143\times143$  &  50-2000   \\
$217\times217$  &  500-2500  \\
$143\times217$  &  500-2500  \\
\hline
\end{tabular}
\end{center}\caption{~\label{tab:cross-spectra} The multipole ranges of the spectra and the cross-spectra provided by Planck data for $\ell \ge 50$ \citep{Planck2013XV}.}
\end{table}

\subsection{Nonparametric angular power spectrum}~\label{subsec:nonparam-spectrum}

We employ the model-independent regression method that was adopted with improvements in \cite{Aghamousa2012,Aghamousa2014}.
In this formalism, a nonparametric fit is
characterized by its \textit{effective degrees of freedom (EDoF)} which
can be considered as the equivalent of the number of
parameters in a parametric regression problem.

The \textit{full-freedom fit} is obtained by minimizing the risk
function under a monotonicity constraints on the shrinkage
parameters \citep{Aghamousa2012}.
The full-freedom fit can be quite
oscillatory depending on the level of noise in the data.
Although this fit is a reasonable fit, captures the essential trend in the data well, and is optimal (in the sense of minimizing a risk function with minimal restrictions), all the
cosmological models predict a far smoother angular power spectrum.
To account for this, we minimize the risk but progressively restrict the EDoF relative to
the full-freedom fit until we get an acceptable smooth fit.
We call this
the \textit{restricted-freedom fit}.
Additional details about this
methodology can be found in \cite{Aghamousa2012}.

\begin{figure}
 \includegraphics[width=\textwidth]{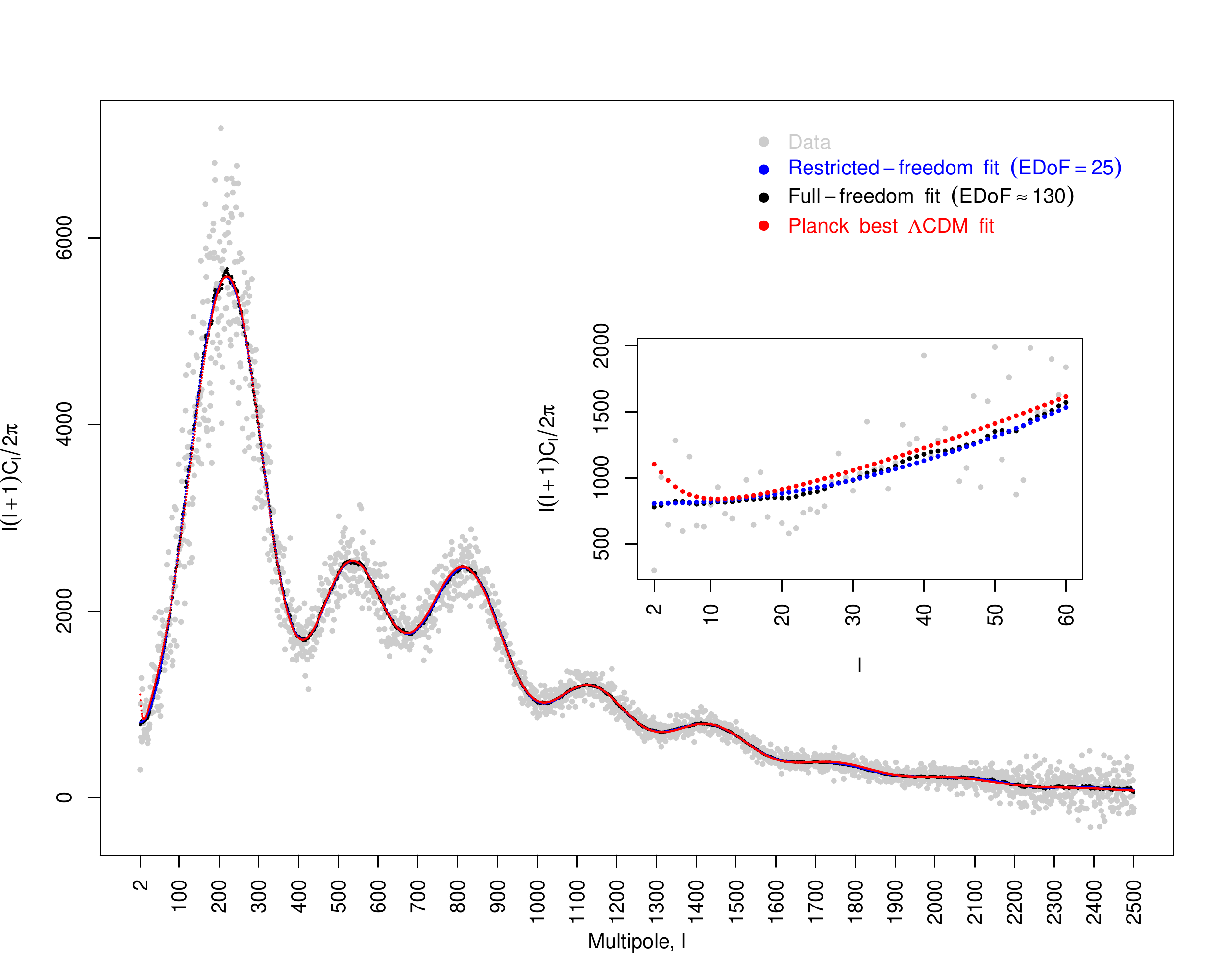}
 \caption{\label{fig:power-spectra1} The nonparametric fits for Planck temperature angular power spectrum data. Black: full-freedom fit (EDoF $\sim 130$), blue: restricted-freedom fit (EDoF $= 25$), red: best-fit $\Lambda$CDM model, gray: Planck angular power spectrum data. The inset plot illustrates the difference between nonparametric fits and the best-fit $\Lambda$CDM model in $\ell<60$ region. The best-fit $\Lambda$CDM model shows an upturn at $\ell<10$ due to ISW effect which is not indicated by the nonparametric reconstruction.}
\end{figure}

Applying this nonparametric method on Planck CMB angular power spectrum data (as described in subsection~\ref{subsec:data}), the full-freedom fit obtained corresponds to EDoF $\sim130$ (Figure~\ref{fig:power-spectra1}; black points).
Although the full-freedom fit does reveal the angular power spectrum, it shows low-level oscillations that arise partly due to the noise in the data.
By restricting the EDoF of the fit, we find that the restricted-freedom fit with EDoF $=25$ (Figure~\ref{fig:power-spectra1}; blue points) has an acceptable level of smoothness.
The choice of EDoF $=25$ is guided by the cosmological consideration that we expect to see a smooth fit with $6$ peaks up to $\ell \sim2000$.
Essentially, by restricting the EDoF, we have attempted to avoid artificial numerical peaks in the fit, and at the same time we try to avoid over-smoothed fits.
At EDoF $=25$, the angular power spectrum clearly reveals six peaks and five dips for $\ell<1850$.
Due to higher noise level for $\ell \geq1850$, the restricted-freedom fit does not resolve individual peaks in this region.
For comparison, we also plot the Planck best-fit $\Lambda$CDM model (Figure~\ref{fig:power-spectra1}; red points).
The full-freedom fit, the restricted-freedom (EDoF $=25$) fit, and the Planck best-fit $\Lambda$CDM by and large follow each other except for one significant difference:
The best-fit $\Lambda$CDM model in Figure~\ref{fig:power-spectra1} (inset plot) shows an up-turn for $\ell<10$; this is expected on the basis of the integrated Sachs-Wolfe effect \citep{Sachs_Wolfe}.
However, neither the full-freedom fit nor the restricted-freedom nonparametric fits, which are driven more by the data than by a model, reveal such an upturn.
We discuss this further in a latter section.

Another useful feature of this nonparametric methodology is a powerful construct called the $(1-\alpha)$-confidence set which, under a minimal set of assumptions, is guaranteed to capture the true but unknown spectrum with probability $(1-\alpha)$ in the limit of large data. This construct can be used to simultaneously quantify uncertainties on any number of features of the fit at the same level of confidence, and also to validate fits based on cosmological models against data.
We will use both aspects of this construct in the next section.

\section{Inferences}~\label{sec:results_discussions}
In this section we present a set of interesting inferences from our nonparametric analysis of the Planck temperature angular power spectrum data.
We use the restricted-freedom nonparametric fit and the corresponding confidence ball for estimating uncertainties related to peaks and dips, for assessing the quality of our nonparametric fit itself, for estimating the acoustic scale $l_A$, and to demonstrate the harmonicity in the peaks.
Further, we employ the full-freedom fit and its confidence ball for testing the consistency of the Planck best-fit $\Lambda$CDM model with the data.

\subsection{Assessment of quality of the nonparametric angular power spectrum}
\label{subsec:quality}

\begin{figure}
 \includegraphics[width=\textwidth]{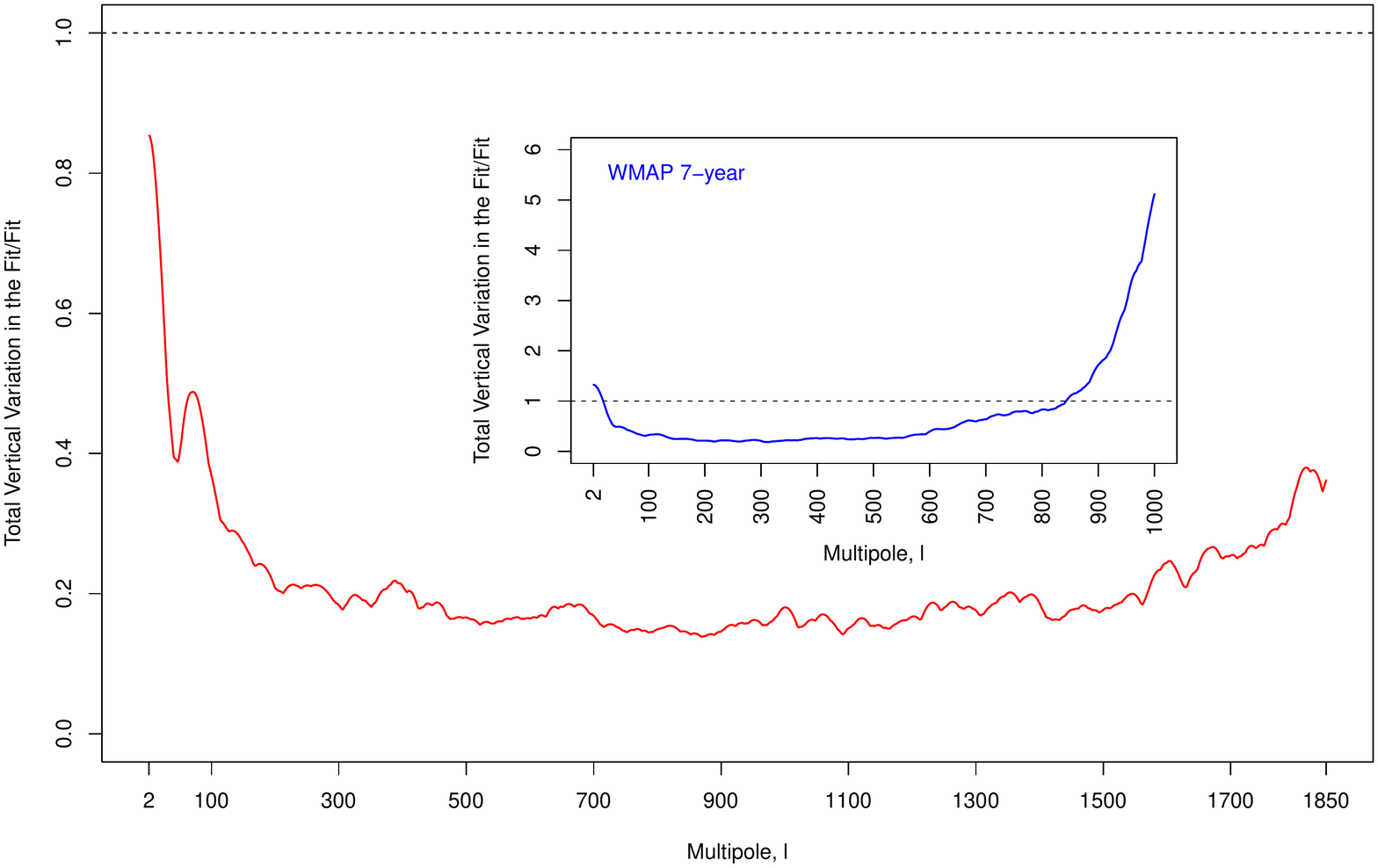}
 \caption{\label{fig:box-car} The total vertical variation of the restricted-freedom fit (EDoF $= 25$) at each $\ell$ within the 95\% confidence ball, divided by the value of the fit. The values less than one indicate the fit is firmly determined by the data. The red curve shows the corresponding ratio for Planck data. It appears that the ratio values are less than one in the whole $\ell<1850$ range. In comparison the similar plot for WMAP 7-year data (inset plot; blue curve) shows the ratios less than 1 at $\ell < 842$. This plot reflects how Planck has improved the quality of the observed CMB data in comparison with WMAP.}
\end{figure}

To assess the quality of our nonparametric fit,
we record the maximum vertical variation in our estimated nonparametric angular power spectrum (restricted-freedom fit with EDoF $=25$) at each multipole using 5000 randomly sampled spectra drawn uniformly from the 95\% confidence ball around the fit.
The ratio of this variation to the value of the fit yields a relative measure of how well the angular power spectrum is determined \citep{GMN+2004,Aghamousa2012}.
We plot (Figure~\ref{fig:box-car}) this ratio over the multipole range of well-resolved peaks ($\ell<1850$).
This ratio has a value less than 1 over this entire range, which indicates that the nonparametric fit is well-determined over this multipole range.
For comparison, we have included (inset) a similar plot for the WMAP 7-year data from \citep{Aghamousa2012}. Here, the ratio is considerably larger than $1$ for $\ell>842$; this is the effect of high noise at high multipoles in that data set.

\subsection{Uncertainties on peaks and dips}~\label{subsec:boxes}

In the world of cosmological models, the shape of the CMB angular power spectrum is dictated by the underlying cosmological model and the parameters therein.
In particular, the locations and heights of peaks and dips in the angular power spectrum are known to depend sensitively on the model parameters.
Consequently, given a cosmological model, the uncertainties in the locations and heights of peaks and dips can in principle be used to assess uncertainties in the values of the cosmological parameters in the model.

\begin{figure}
 \includegraphics[width=\textwidth]{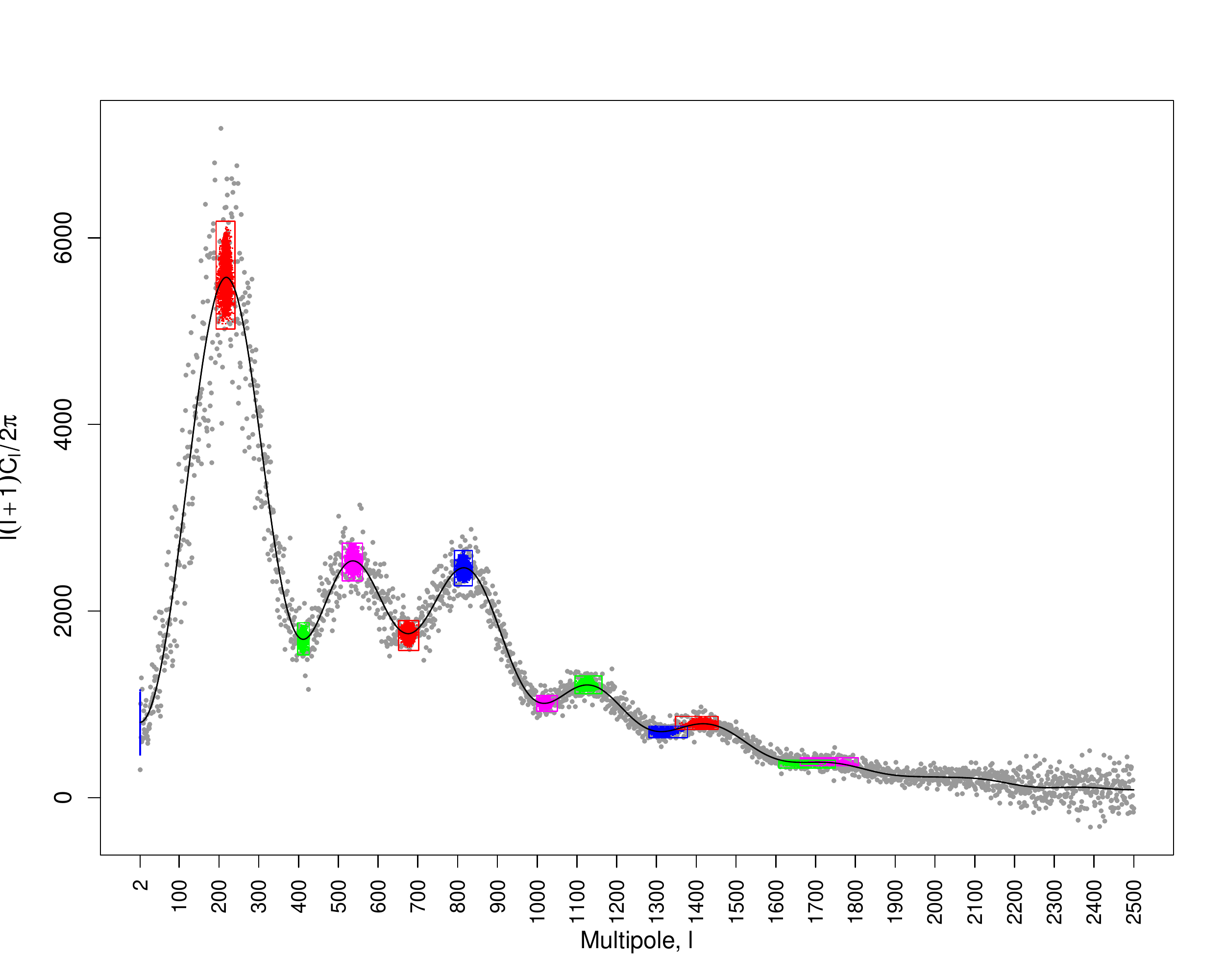}
 \caption{\label{fig:uncertainty-boxes} The 95\% uncertainty boxes around the peaks and dips. The black curve is the restricted-freedom fit with EDoF $=25$ to the Planck angular power spectrum data (gray points). The uncertainty boxes shows the extreme variations of 5000 acceptable sampled angular power spectra from the 95\% confidence ball. The vertical blue line at $\ell=2$ indicates the 95\% variation of the fit and does not represent the uncertainty of a peak or dip.}
\end{figure}

The nonparametric methodology used in this paper allows computing uncertainties (in the form of confidence intervals) on any feature of the fit using the confidence ball around the fit. Specifically, for this purpose, we apply the prescription in \citep{Aghamousa2012} to the six peaks and five dips in the $\ell < 1850$ multipole region using 5000 random samples drawn uniformly from the 95\% confidence ball around the restricted-freedom fit with EDoF$=25$. The most extreme variation in the peaks and dips, both horizontal and vertical, constitute the corresponding 95\% confidence intervals; these are shown in Fig.\ \ref{fig:uncertainty-boxes} as rectangles around the peaks and dips.
The restricted-freedom fit does not show peak or dip in multipoles $\ell<10$.
Therefore the vertical blue line in Figure~\ref{fig:uncertainty-boxes} at $\ell=2$ indicates only the 95\% variation of the fit.
These confidence intervals are tabulated in Table~\ref{tab:boxes}.

\begin{table}
 \begin{tabular}{c|c|c|c}
  Peak Location &  Peak Height & Dip Location & Dip Height \\
  \hline

   $\ell_1: \hfill  (193, 240)$ & $h_1: \hfill (5020.876, 6176.201)$ & $\ell_{1+{1 \over 2}}: \hfill  (398, 427)$ & $h_{1+{1 \over 2}}: \hfill  (1529.765, 1874.690) $  \\
   $\ell_2: \hfill  (510, 561)$ & $h_2: \hfill (2322.606, 2726.660)$ & $\ell_{2+{1 \over 2}}: \hfill  (652, 702)$ & $h_{2+{1 \over 2}}: \hfill  (1577.765, 1898.944) $ \\
   $\ell_3: \hfill  (792, 837)$ & $h_3: \hfill (2270.030, 2649.854)$ & $\ell_{3+{1 \over 2}}: \hfill (999, 1051)$ & $h_{3+{1 \over 2}}: \hfill  (927.889, 1095.580) $ \\
   $\ell_4: \hfill (1095, 1163)$ & $h_4: \hfill (1113.821, 1302.823)$ & $\ell_{4+{1 \over 2}}: \hfill (1281, 1378)$ & $h_{4+{1 \over 2}}: \hfill  (643.499, 766.079) $  \\
   $\ell_5: \hfill (1348, 1455)$ & $h_5: \hfill (731.262, 872.029)$ & $\ell_{5+{1 \over 2}}: \hfill (1608, 1750$ & $h_{5+{1 \over 2}}: \hfill  (318.291, 402.112) $  \\
   $\ell_6: \hfill (1661, 1808)$ & $h_6: \hfill (342.363, 428.616)$ &                   ...
                &                 ...
                       \\
 \end{tabular}

 \caption{\label{tab:boxes} The nonparametric 95\% confidence intervals on location and height of peaks and dips of the restricted-freedom fit (EDoF $=25$).
Here, $\ell_m$ ($h_m$) stands for the location (height) of the $m$th peak, and $\ell_{m+{1 \over 2}}$ ($h_{m+{1 \over 2}}$) denotes the location (depth) of the $m$th dip.}
\end{table}

\subsection{Harmonicity of peaks and the acoustic scale}~\label{subsec:harmonicity_acoustic}

In the early Universe before the last scattering, the baryon-photon fluid underwent acoustic oscillations.
The imprint of these oscillations is expected to be a series of harmonic peaks in the angular power spectrum.
From a model-independent viewpoint, it would be interesting to see if this harmonicity can be ``rediscovered'' in a nonparametric analysis such as ours.

\begin{figure}
 \includegraphics[width=\textwidth]{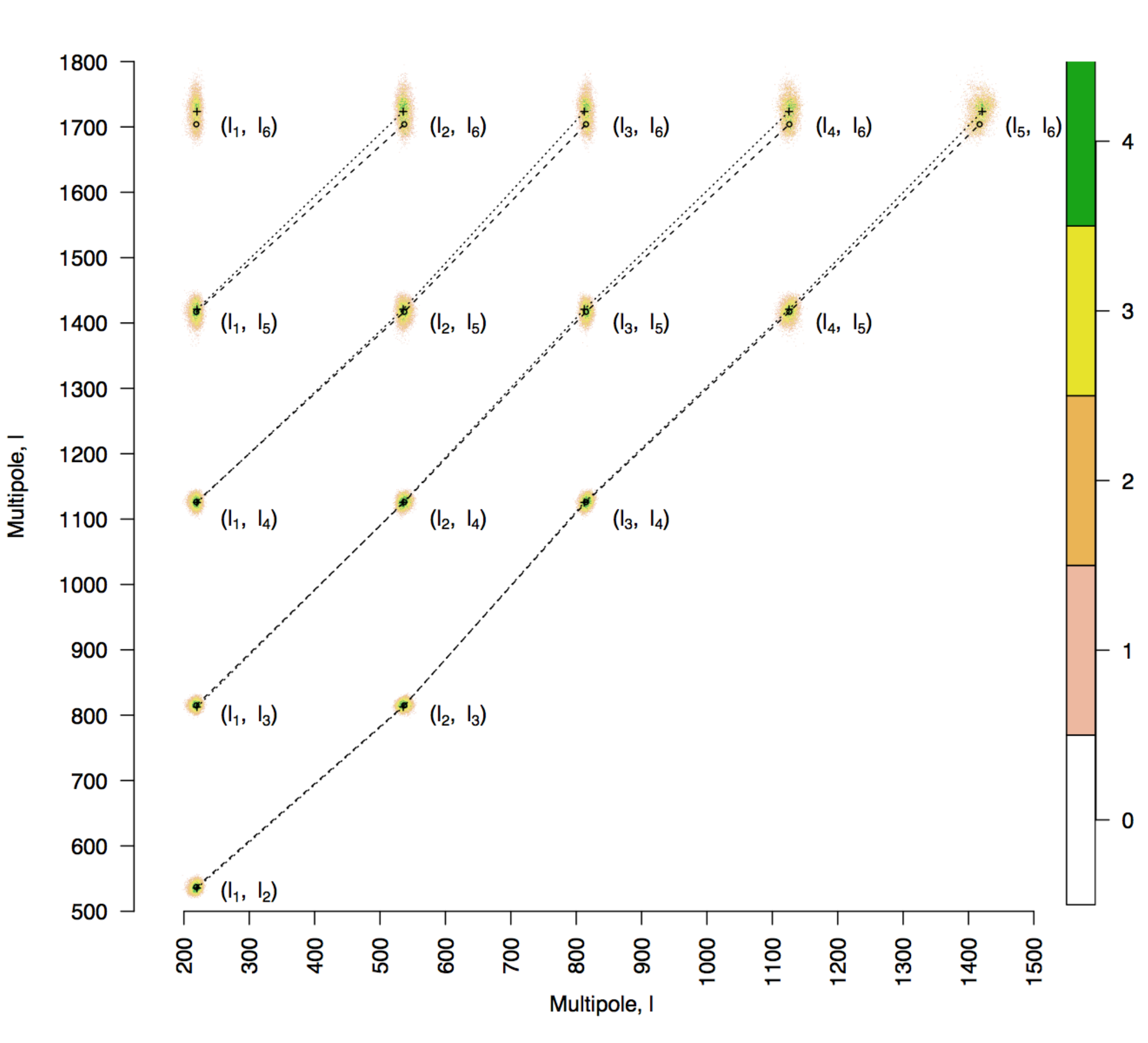}
 \caption{\label{fig:harmonicity} Nonparametric indication of the harmonicity of peaks in the CMB angular power spectrum. This is a 2-dimensional color-coded histogram for the $(\ell_i,\ell_j)$ scatter, where $\ell_i$ is the position of the $i$th peak, for a sample of 10000, 6-peaked spectra from the 95\% confidence set. Color indicates log(count) over an $\ell$-grid with unit spacing in both directions. The peak-peak locations for the $\Lambda$CDM parametric fit \citep{Planck2013XVI, Planck2013XV} (crosses), and those for our nonparametric fit (circles) are connected by dotted and dashed lines respectively. We interpret the arrangement of these peaks as nonparametric evidence for the basic physics of harmonicity of acoustic oscillations of the baryon-photon fluid that gave rise to CMB anisotropies.}
\end{figure}

Our results in this regard are illustrated in Figure~\ref{fig:harmonicity}.
This figure is the result of a random sample of 10000 6-peaked spectra uniformly sampled from the 95\% confidence set around the restricted-freedom fit, and shows a 2-dimensional color-coded histogram for the $(\ell_i,\ell_j)$ scatter, where $\ell_i$ is the position of the $i$th peak.
Also indicated in this figure are the peak-against-peak positions for two single fits; namely, the angular power spectrum for the best-fit $\Lambda$CDM model \citep{Planck2013XVI, Planck2013XV} (crosses), and our nonparametric restricted-freedom fit (circles).

Two features of this remarkable plot are worth pointing out. First, we see a close agreement between our nonparametric fit and the $\Lambda$CDM-based parametric fit.
Second, we see a regular lattice structure in the (nonparametric) peak-against-peak scatter.
We interpret this latter as a direct, model-independent evidence for the basic physics of harmonicity of acoustic oscillations in the baryon-photon fluid.

\begin{figure}
 \includegraphics[width=\textwidth]{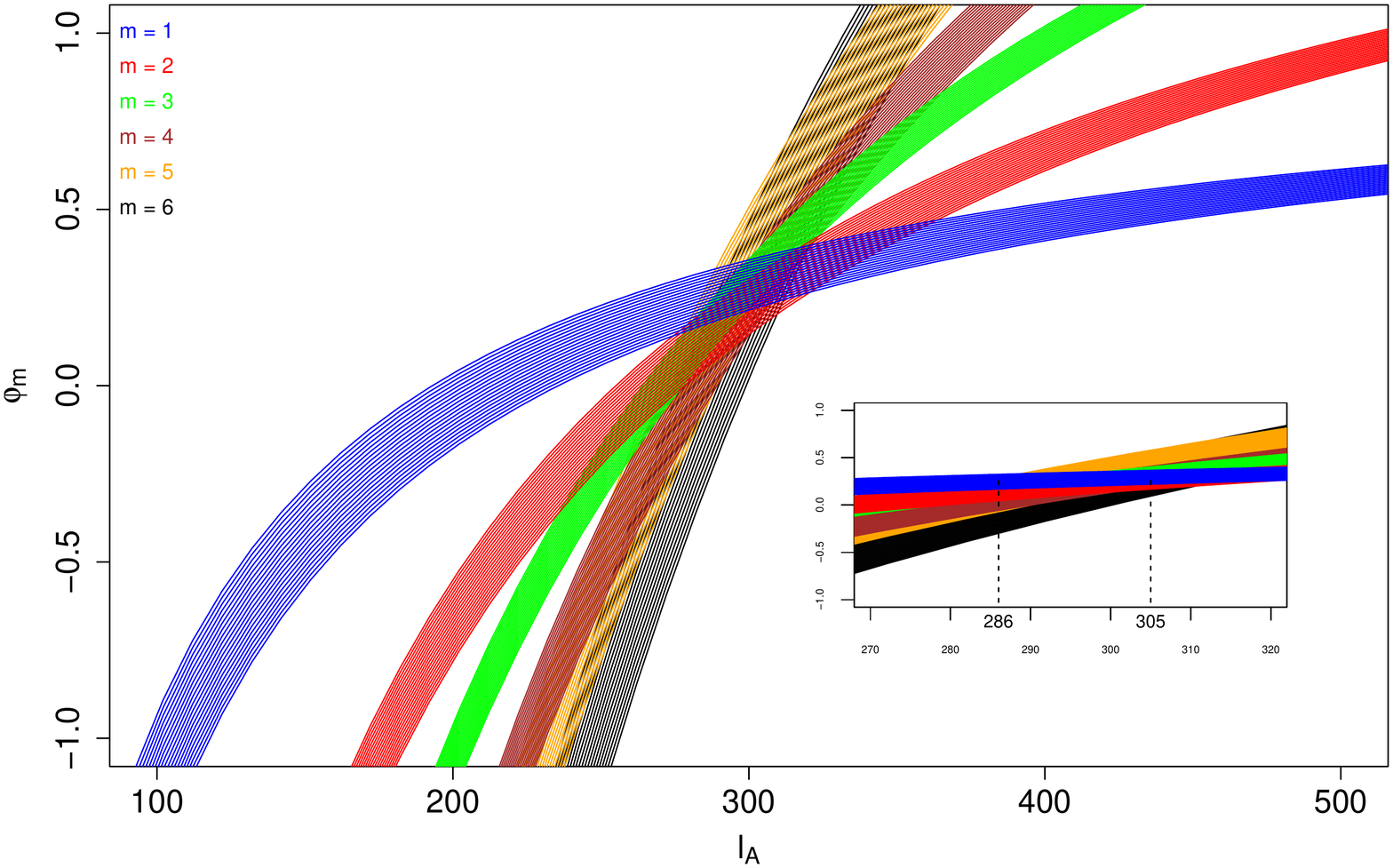}
 \caption{\label{fig:acoustic} The 95\% confidence bands of phase shift $\phi_m$ versus acoustic scale $\ell_{A}$ as derived from the 95\% confidence intervals of first six peak location of the restricted-freedom fit (EDoF $=25$). The intersection suggests the acoustic scale value $286 \leq \ell_{A} \leq 305$.}
\end{figure}

\paragraph{The acoustic scale $l_A$.}
The harmonicity of acoustic oscillations can also be stated in terms of the acoustic scale $\ell_A$.
The equation $\ell_{A}=m(\ell_m-\phi_m)$ expresses the relation between location $\ell_m$ of $m$th peak, the acoustic scale $\ell_{A}$ and the phase shift parameter $\phi_m$ \citep{HFZ+2001, DL2002}.
This relationship, together with the 95\% confidence intervals (Table~\ref{tab:boxes}) for the locations of the first six peaks
lead to hyperbolic confidence bands in the $\ell_{A} - \phi_m$ plane (Figure~\ref{fig:acoustic}).
A physically meaningful range of values for $\phi_m$ is $\vert \phi_m \vert < 1$.
The intersection of these bands (which occurs, incidentally, for $\vert \phi_m \vert < 1$) determine an estimated 95\%
confidence interval for the acoustic scale, namely, $286 \leq \ell_{A} \leq 305$.

\subsection{Model-checking by data}~\label{subsec:checking}

The high-dimensional confidence ball around the fit can also be employed to test the extent to which different cosmological models are supported by the data, as follows.
We measure the distance of an angular power spectrum predicted by a specific cosmological model from the center of the confidence ball.
The confidence level corresponding to this distance can be interpreted as the probability of rejection of the  cosmological model under consideration (for details refer to \cite{GMN+2004,Aghamousa2012}).
It is worth noting that for providing the background angular power spectrum data, we have used the foregrounds nuisance parameters values associated with the Planck best-fit $\Lambda$CDM model (subsection~\ref{subsec:data}).
Under these circumstances this formalism can be used specially to check the consistency of the Planck best-fit $\Lambda$CDM model with the data.
The confidence distance of the best-fit $\Lambda$CDM angular power spectrum turns out to be about 36\% from the center of the confidence ball (i.e., our restricted-freedom nonparametric fit), implying a 36\% probability of being rejected as a candidate for the true but unknown spectrum.
In comparison, our earlier analysis of the WMAP 7-year data found the best-fit $\Lambda$CDM angular power spectrum \citep{Aghamousa2012} at 9\% confidence distance from the corresponding nonparametric fit.


Another assessment of a cosmological model can be performed by using the $\chi^2$ value of associated angular power spectrum.
In our analysis, we used the Planck likelihood code \citep{Planck2013XV} to calculate the $\chi^2$ value of the Planck best $\Lambda$CDM fit and a number of restricted-freedom fits.
Since the restricted-freedom fits with small values of EDoF are physically meaningless, we consider  restricted-freedom fits with $11\leq \text{EDoF} \leq130$ (130 is the EDoF for the full-freedom fit).

\begin{figure}
 \includegraphics[width=\textwidth]{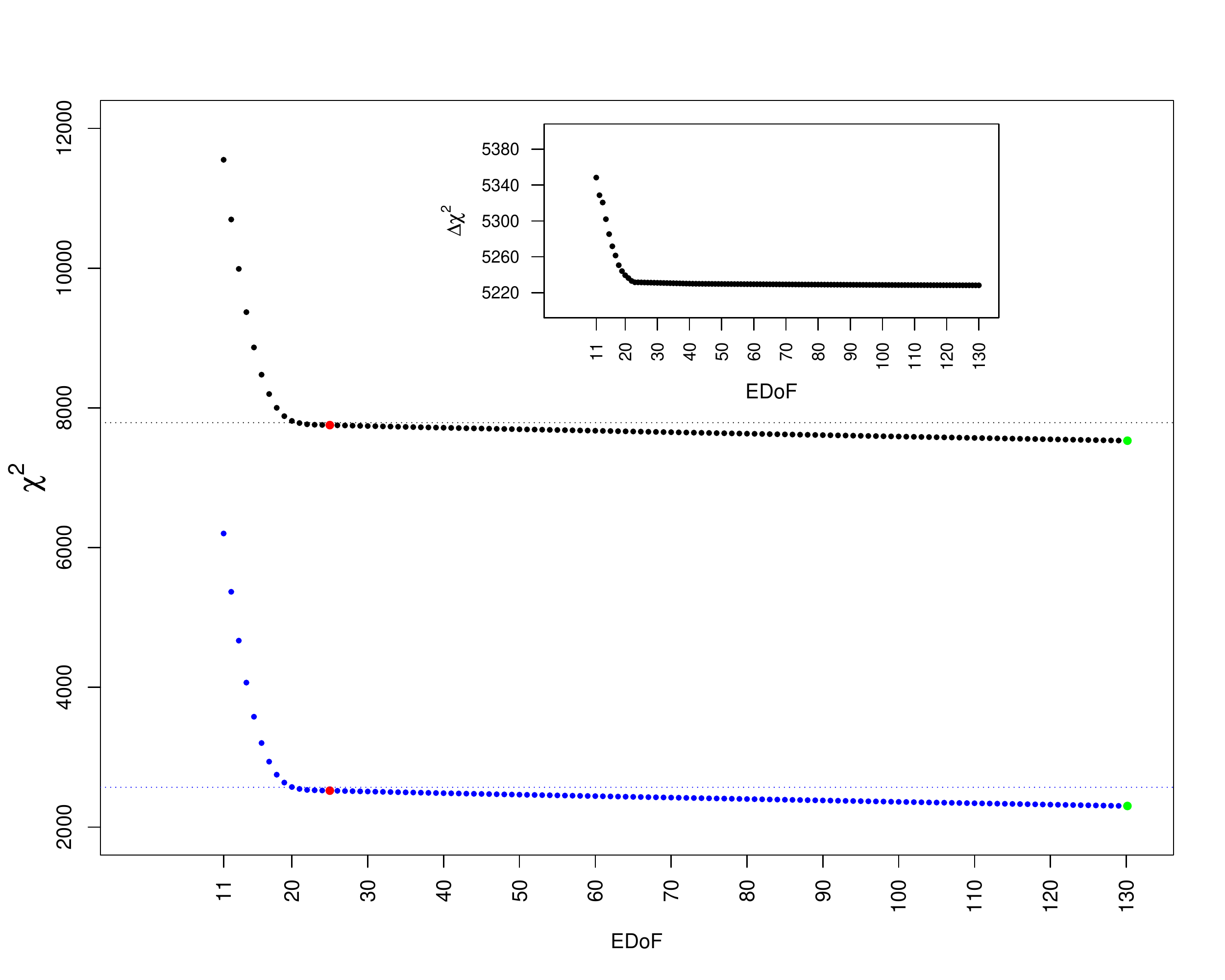}
 \caption{\label{fig:chi2_1} The $\chi^2$ values for nonparametric fits computed by the Planck likelihood code (black points) and the calculated covariance matrix (blue points) in subsection~\ref{subsec:data} (Equation~\ref{equ:weight-covariance}). Both $\chi^2$ values follow the similar trend with an off-set difference. The $\chi^2$ values of the Planck best-fit $\Lambda$CDM model by likelihood code and covariance matrix are drawn by black and blue dotted horizontal lines respectively. Furthermore the $\chi^2$ values of the full-freedom fit (EDoF $\sim130$) and the restricted-freedom fit (EDoF $=25$) are marked by green and red colors respectively. The off-set value $(\Delta \chi^2)$ is plotted in inset graph which shows a nearly constant $\Delta \chi^2$.}
\end{figure}

Alternatively, we calculate the $\chi^2$ values of these fits by using the calculated covariance matrix in subsection~\ref{subsec:data} (Equation~\ref{equ:weight-covariance}) as following,
  \begin{equation}
  \label{eq:chi2}
\chi^2= (\widehat{{\cal D}}_{\ell} - \bar{{\cal D}_{\ell}})^T \Sigma^{-1} (\widehat{{\cal D}}_{\ell} - \bar{{\cal D}_{\ell}})
 \end{equation}
where $\widehat{{\cal D}}_{\ell}$ is the angular power spectrum fit, $\bar{{\cal D}_{\ell}}$ is the weighted-average angular power spectrum data with covariance matrix $\Sigma=\text{Cov}(\bar{{\cal D}_{\ell}},\bar{{\cal D}_{\ell'}})$ (obtained in Subsection~\ref{subsec:data}).
Figure~\ref{fig:chi2_1} illustrates the $\chi^2$ values computed by the Planck likelihood code (black points) and by the covariance matrix (blue points).
For comparison, we also indicate the $\chi^2$ values of the Planck best-fit $\Lambda$CDM model computed using the likelihood code and covariance matrix approaches with black and blue horizontal lines respectively.
In addition, the $\chi^2$ values of the full-freedom fit (EDoF $\sim130$) and the restricted-freedom fit (EDoF $=25$) are marked by green and red colors respectively on the corresponding $\chi^2$ curve.

We see that the $\chi^2$ values computed using the covariance matrix approach follow approximately the ones computed using the Planck likelihood code, except for an offset.
The difference between two $\chi^2$ values (Figure~\ref{fig:chi2_1}, inset graph) suggests that this offset is nearly constant for $\ell \gtrsim 25$.
This suggests that the derived covariance matrix we used in this work can be used for a reasonable approximation of the likelihood to the Planck data. 
In fact, one could take this analysis further by deriving the foreground parameters iteratively after each round of angular power spectrum reconstruction, where the foreground parameters from the best-fit $\Lambda$CDM model are used only as an initial guess which the method could improve upon iteratively.
However, this is beyond the scope of the present work.

\section{Conclusion}~\label{sec:conclusions}

In this work, we have presented a nonparametric analysis of the Planck CMB temperature angular power spectrum data in order to estimate the form and characteristics of the angular power spectrum, using a nonparametric methodology \citep{Beran1996, BD1998, Beran2000, Beran2000b, GMN+2004,BSM+2007,Aghamousa2012}.
The main goal of this paper is to make a comparison between a nonparametric estimate of the angular power spectrum and the parametric Planck best-fit $\Lambda$CDM spectrum.
We have therefore used the nuisance parameters associated with the Planck best-fit $\Lambda$CDM model to obtain the background angular power spectrum data. 
At multipoles $\ell \ge50$, we have calculated the weighted-average of the Planck angular power spectra in the frequency channels $100 \times 100$ GHz, $143\times143$ GHz, $217\times217$ GHz and $143\times217$ GHz.
We have also calculated the covariance matrix of the weighted-angular power spectrum data by taking into account all the correlation terms between spectra.

Our nonparametric angular power spectrum resolves the peaks and dips up to $\ell \sim1850$ and we have shown that the quality of estimated angular power spectrum is reasonably acceptable.
The small 95\% confidence intervals of locations and heights of peaks and dips reflect the accuracy of the estimation.
These results lead to a nonparametric demonstration of the harmonicity of peaks of angular power spectrum (and acoustic scale $\ell_{A}$) where we can compare them with the predictions of the concordance model of cosmology.
We also check the consistency of the Planck best-fit $\Lambda$CDM model with the data.
$\Lambda$CDM model seems to be at $36\%$ confidence distance ($36\%$ chance that model is wrong) from the center of confidence ball which is substantially further than the $9\%$ distance derived earlier analyzing WMAP 7-year data by \cite{Aghamousa2012}.
This hints that the data may suggest some unexpected features beyond the flexibility of the standard model but we need more data to make better evaluation \citep{Hazra2014a, Hazra2014b, Larson2014}.
We should also note that our direct nonparametric reconstruction of the CMB angular power spectrum indicates no upturn at low multiples which is not exactly what we expect from ISW effect.
However, due to large uncertainties at low multiples due to cosmic variance it is not possible to address this issue more quantitatively.

\section*{Acknowledgments}
AA and AS wish to acknowledge support from the Korea Ministry
of Education, Science and Technology, Gyeongsangbuk-Do and Pohang City for Independent Junior Research Groups at the Asia Pacific Center for Theoretical Physics.
The authors would like to thank Simon Prunet and Dhiraj Hazra for useful discussions.
AS would like to acknowledge the support of the National Research Foundation of Korea (NRF-2013R1A1A2013795).
We acknowledge the use of Planck data and likelihood from Planck Legacy Archive (PLA).

\bibliographystyle{JHEP}
\bibliography{/home/amir/Desktop/amir16_backup/Post-Doc1/reference}

\providecommand{\href}[2]{#2}\begingroup\raggedright\begin{thebibliography}{10}

\bibitem{Aghamousa2012}
A.~Aghamousa, M.~Arjunwadkar, and T.~Souradeep, {\it Evolution of the cosmic
  microwave background power spectrum across wilkinson microwave anisotropy
  probe data releases: A nonparametric analysis},  {\em Astrophys. J.} {\bf
  745} (2012), no.~2 114.

\bibitem{Beran1996}
R.~Beran, {\it Confidence sets centred at $c_p$-estimators},  {\em Ann.\ Inst.\
  Statist.\ Math.} {\bf 48} (1996) 1--15.

\bibitem{BD1998}
R.~Beran and L.~D\"umbgen, {\it Modulation of estimators and confidence sets},
  {\em Ann.\ Statist.} {\bf 26} (1998), no.~5 1826--1856.

\bibitem{Beran2000}
R.~Beran, {\it React scatterplot smoothers: Superefficiency through basis
  economy},  {\em J.\ Amer.\ Statist.\ Assoc.} {\bf 95} (2000), no.~449
  155--171.

\bibitem{Beran2000b}
R.~Beran, {\it React trend estimation in correlated noise},  in {\em
  Asymptotics in statistics and probability: papers in honor of George Gregory
  Roussas} (M.~L. Puri, ed.), pp.~1--16.
\newblock VSP International Science Publishers, 2000.

\bibitem{GMN+2004}
C.~R. Genovese, C.~J. Miller, R.~C. Nichol, M.~Arjunwadkar, and L.~Wasserman,
  {\it Nonparametric inference for the cosmic microwave background},  {\em
  Statist. Sci.} {\bf 19} (2004), no.~2 308--321.

\bibitem{BSM+2007}
B.~Bryan, J.~Schneider, C.~J. Miller, R.~C. Nichol, C.~R. Genovese, and
  L.~Wasserman, {\it Mapping the cosmological confidence ball surface},  {\em
  Astrophys. J.} {\bf 665} (August, 2007) 25--41.

\bibitem{Aghamousa2014}
A.~Aghamousa, M.~Arjunwadkar, and T.~Souradeep, {\it Model-independent
  forecasts of cmb angular power spectra for the planck mission},  {\em Phys.
  Rev. D} {\bf 89} (Jan, 2014) 023509.

\bibitem{TPC2006}
{The Planck Collaboration}, {\it {The Scientific Programme of Planck}},  {\em
  ArXiv Astrophysics e-prints} {\bf astro-ph/0604069} (Apr., 2006).

\bibitem{Planck2013I}
P. A. R. Ade et al. {\it Planck 2013 results. I overview of products and scientific results}, {\em A\&A} {\bf 571} (2014) A1.

\bibitem{Planck2013XV}
P. A. R. Ade et al. {\it Planck 2013 results. XV cmb power spectra and likelihood}, {\em A\&A} {\bf 571} (2014) A15.

\bibitem{Planck2013XVI}
P. A. R. Ade et al. {\it Planck 2013 results. XVI cosmological parameters},  {\em A\&A} {\bf 571} (2014) A16.

\bibitem{Sachs_Wolfe}
R.~K. {Sachs} and A.~M. {Wolfe}, {\it {Perturbations of a Cosmological Model
  and Angular Variations of the Microwave Background}},  {\em Astrophys. J.}
  {\bf 147} (Jan., 1967) 73.

\bibitem{HFZ+2001}
W.~{Hu}, M.~{Fukugita}, M.~{Zaldarriaga}, and M.~{Tegmark}, {\it {Cosmic
  Microwave Background Observables and Their Cosmological Implications}},  {\em
  Astrophys. J.} {\bf 549} (Mar., 2001) 669--680.

\bibitem{DL2002}
M.~Doran and M.~Lilley, {\it The location of cmb peaks in a universe with dark
  energy},  {\em Mon. Not. Roy. Astron. Soc.} {\bf 330} (2002) 965.

\bibitem{Hazra2014a}
D.~K. {Hazra} and A.~{Shafieloo}, {\it {Confronting the concordance model of
  cosmology with Planck data}},  {\em J. Cosmology Astropart. Phys.} {\bf 1}
  (Jan., 2014) 43, [\href{http://xxx.lanl.gov/abs/1401.0595}{{\tt
  arXiv:1401.0595}}].

\bibitem{Hazra2014b}
D.~K. {Hazra} and A.~{Shafieloo}, {\it {Test of consistency between Planck and
  WMAP}},  {\em Phys. Rev. D} {\bf 89} (Feb., 2014) 043004,
  [\href{http://xxx.lanl.gov/abs/1308.2911}{{\tt arXiv:1308.2911}}].

\bibitem{Larson2014}
D.~{Larson}, J.~L. {Weiland}, G.~{Hinshaw}, and C.~L. {Bennett}, {\it
  {Comparing Planck and WMAP: Maps, Spectra, and Parameters}},  {\em ArXiv
  e-prints} (Sept., 2014) [\href{http://xxx.lanl.gov/abs/1409.7718}{{\tt
  arXiv:1409.7718}}].


\end{thebibliography}\endgroup

\end{document}